\documentclass[aps,prl,twocolumn,preprintnumbers,amsmath,amssymb]{revtex4}
\usepackage{graphicx}      % alternative graphics specifications
\usepackage{times}
\usepackage[colorlinks=true,citecolor=blue]{hyperref}

\newcommand{\ra}{\rangle}

\newcommand{\mcal}{\mathcal}
\newcommand{\f}{\frac}

\newcommand{\te}{\text}
\newcommand{\tb}{\textbf}
\newcommand{\omg}{\omega}

\newcommand{\Omg}{\Omega}

\newcommand{\var}{\varepsilon}
\newcommand{\pa}{\partial}

\begin{document}
    %\linenumbers
    \title{Enhanced Two-Photon Processes in Quantum Dots inside Photonic Crystal Nanocavities and Quantum Information Science Applications}
    \author{Ziliang Lin}
        \email{carterlin@stanford.edu}
    \author{Jelena Vu\v{c}kovi\'{c}}
    \affiliation{E. L. Ginzton Laboratory, Stanford University, Stanford, CA 94305, USA}
    \date{\today}
\begin{abstract}
We show that the two-photon transition rates of quantum dots coupled
to nanocavities are enhanced by up to $8$ orders of magnitude
relative to quantum dots in bulk host. We then propose how to take
advantage of this enhancement to implement coherent quantum dot
excitation by two-photon absorption, entangled photon pair
generation by two-photon spontaneous emission, and single-photon
generation at telecom wavelengths by two-photon stimulated and
spontaneous emission.
\end{abstract}

\maketitle

%\section{Introduction}

Recent progress in semiconductor quantum dots (QDs) coupled to
photonic crystal (PC) cavities has shown that these systems exhibit
cavity quantum electrodynamics effects. Experiments have
demonstrated creation of single photons in the near
infrared\cite{article:eng07b}, Purcell enhancement of QD spontaneous
emission rates in weak coupling
regime\cite{article:lau05,article:eng05}, polariton state splitting
in strong coupling
regime\cite{article:hen07,article:yos04,article:eng07}, controlled
amplitude and phase shifts at a single photon
level\cite{article:fus08}, and photon induced tunneling and
blockade\cite{article:far08}. These results indicate that the QDs
can be combined with PC cavities in an integrated platform for
quantum information
science\cite{article:cir97,article:dua01,article:dua04}.

However, many important challenges remain in the QD-PC cavity
systems. Among them is the generation of indistinguishable single
photons, preferably at telecom wavelengths. This challenge results
from the fact that an incoherently excited QD has to undergo phonon
assisted relaxation to its lowest excited state, so the single
photons it emits have different temporal
profiles\cite{article:vuc06,article:bec02}. For this reason, the
maximum demonstrated indistinguishability between photons emitted
from a single QD has been around $81\%$\cite{article:san02}. This
problem would be resolved by exciting QDs coherently. In addition,
coherent excitation is of great interest for quantum information
science, as it also enables manipulation of QD states. However, this
type of excitation experiments face the difficulty of separating
signal and probe photons that have the same wavelength. Another
challenge is the generation of entangled photon
pairs\cite{article:ben00}. Although photon pairs can be created from
biexciton decay cascade in a QD\cite{article:ako06,article:ste05},
the pairs are not entangled in polarization because the intermediate
exciton states are split by anisotropic electron-hole exchange
interactions\cite{article:gam96,article:bay02}.

\begin{figure}
\includegraphics[scale=.38]{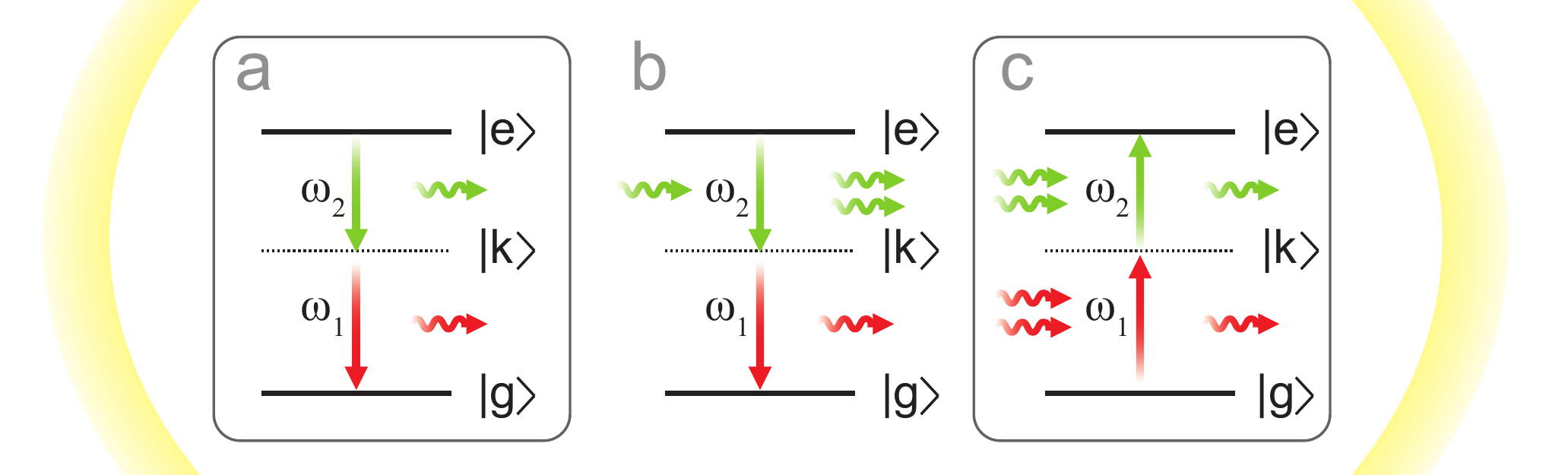}
\caption{\label{fig:two_photon} (Color online) (a) Spontaneous
emission of a pair of photons from an excited QD via a virtual
state. (b) Stimulated two-photon emission from an excited QD. The
first photon emission is stimulated with an external laser and the
second photon emission is spontaneous. (c) Two-photon absorption by
a QD. In all three processes, the transition rates are enhanced
inside a single-mode or double-mode cavity (pictured here as a
generic cavity) matching the photon frequencies.}
\end{figure}

In this Letter, we analyze two-photon absorption (TPA) and emission
(TPE) in QDs coupled to PC cavities as solutions to these challenges
(Fig.~\ref{fig:two_photon}). TPA allows coherent excitation of the
QDs when the frequencies of the two photons are tuned such that
their sum matches the transition frequency of the QDs, and at the
same time allows for the frequency separation of the signal and
probe photons. Hence, TPA can eliminate jitter in single photon
generation from a QD. On the other hand, TPE from an excited QD
allows the generation of single photons and entangled photon pairs
on demand at telecom wavelengths. We also show that semiconductor
QDs coupled to PC cavities enable great enhancement of two-photon
transition rates relative to slow rates in bulk
host\cite{article:pad07,article:xin08,article:hay08}.

%\section{Transition Rate Calculations and Discussion}

We model the semiconductor QD as a two-level system with energy
separation $\hbar\omg_d$, where $\hbar$ is the reduced Planck's
constant. The QD is illuminated by two laser beams of frequencies
$\omg_1$ and $\omg_2$. By applying time-dependent perturbation
theory, we obtain the nondegenerate two-photon transition rate from
the ground state to the excited state: $\gamma =
2\pi|\Omg_\te{eff}|^2\delta(\omg_d-\omg_1-\omg_2)$, where the
effective Rabi rate is
\begin{equation}\label{eqn:nondegenerate_transition}
    \Omg_\te{eff}=\sum_k\left[\f{\Omg_{gk,1}\Omg_{ke,2}}{\Delta_{gk,1}}+\f{\Omg_{gk,2}\Omg_{ke,1}}{\Delta_{gk,2}}\right].
\end{equation}
The summation is performed over all the virtual intermediate states
$|k\ra$'s. $\Omg_{gk,1}$ ($\Omg_{gk,2}$) is the Rabi transition rate
between the ground state and the intermediate state driven by the
laser with frequency $\omg_1$ ($\omg_2$). $\Delta_{gk,1}$ is the
QD-light detuning $\omg_k-\omg_g-\omg_1$, and a similar expression
holds for $\Delta_{gk,2}$. The delta function represents energy
conservation. It must be pointed out that although
Eqn.~\eqref{eqn:nondegenerate_transition} describes TPA rates, TPE
rate expression holds the same form with $\Delta_{gk,1}$
($\Delta_{gk,2}$) replaced by $\Delta_{ke,2}$ ($\Delta_{ke,1}$). We
start from the expression for two-photon Rabi frequency for an
excited QD with quantized electric field:
\begin{equation}\label{eqn:Omega}
    \Omg_{gk,1}
    = \f{|\tb d_{gk}|}\hbar\left[\f{(N_1+1)\hbar\omg_1}{2n^2\var_0
    V_1}\right]^{1/2}\psi_{gk,1},
\end{equation}
where $\tb d_{gk}$ is the transition dipole moment. $N_1$ and $V_1$
are photon number and volume of the mode at frequency $\omg_1$ that
the QD is interacting with. $n$ is refractive index of the host
substrate and $\var_0$ is the free space permittivity. $\psi_{gk,1}$
characterizes the reduction in Rabi rate due to position mismatch
between the QD and the electric field maximum within the cavity:
$\psi_{gk,1} = \f{|\tb E_1(\tb r)|}{|\tb E_1(\tb r_M)|}\left(\f{\tb
d_{gk}\cdot\tb e_1}{|\tb d_{gk}|}\right)$, where $\tb E_1(\tb r)$ is
the electric field at the QD location, $|\tb E_1(\tb r_M)|$ is the
maximum field strength in the cavity, and $\tb e_1$ is the
polarization of the field. Similar expressions hold for Rabi rates
$\Omg_{ke,1}$, $\Omg_{gk,2}$, and $\Omg_{ke,2}$.

By combining Eqs.~\eqref{eqn:nondegenerate_transition} and
~\eqref{eqn:Omega}, we can write the expression for nondegenerate
two photon transition rate as:
\begin{equation}\label{eqn:TPSE}
    \gamma = 2\pi\f{\omg_1\omg_2(N_1+1)(N_2+1)}{(2\hbar n^2\var_0)^2V_1V_2}
    M_{12}^2\delta(\omg_d-\omg_1-\omg_2),
\end{equation}
with
\begin{equation*}
    M_{12} = \left|\sum_k\left[\f{|\tb d_{gk}||\tb d_{ke}|\psi_{gk,1}\psi_{ke,2}}{\Delta_{gk,1}}+\f{|\tb d_{gk}||\tb
    d_{ke}|\psi_{gk,2}\psi_{ke,1}}{\Delta_{gk,2}}\right]\right|.
\end{equation*}
In Eqs.~\eqref{eqn:Omega} and~\eqref{eqn:TPSE}, for spontaneous
emission we set $N_1=0$, for stimulated emission we keep $N_1+1$
term, and for absorption we replace $N_1+1$ with $N_1$. To obtain
the total two-photon spontaneous emission (TPSE) rate
$\Gamma^\te{TPSE}$ (Fig.~\ref{fig:two_photon}a), we need to
integrate Eqn.~\eqref{eqn:TPSE} over all frequencies:
$\pa\Gamma^\te{TPSE}/\pa\omg_2=\int d\omg_1\
\gamma\rho(\omg_1)\rho(\omg_2)$, where $\rho$'s are the photon
densities of states. By setting $N_1=N_2=0$, we obtain
\begin{equation*}
    \f{\pa\Gamma^\te{TPSE}}{\pa\omg_2} = \f{2\pi}{(2\hbar n^2\var_0)^2V_1V_2}
    \omg_1\omg_2 \rho(\omg_1)\rho(\omg_2)M_{12}^2.
\end{equation*}

We arrive at our final expression by writing explicitly the density
of states $\rho(\omg)$. In bulk, $\rho(\omg_1) =
V_1n^3\omg_1^2/(3\pi^2c^3)$. For a cavity $\omg_1\rho(\omg_1) =
2Q_1\phi_1/\pi$, where $Q_1$ is the cavity quality factor and
$\phi_1$ characterizes frequency mismatch between $\omg_1$ and
cavity resonance frequencies $\omg_{1c}$, $\phi_1(\omg_1) =
\f{\omg_1/\omg_{1c}}{1+4Q_1^2(\omg_1/\omg_{1c}-1)^2}$. Finally, the
TPSE rates for a QD in bulk semiconductor host
($\Gamma^\te{TPSE}_0(\omg_1,\omg_2)$) and in a double-mode cavity
centered at $\omg_{1c}$ and $\omg_{2c}$
($\Gamma^\te{TPSE}_{\omg_{1c},\omg_{2c}}(\omg_1,\omg_2)$) are
\begin{align}\label{eqn:all}
    \f{\pa\Gamma^\te{TPSE}_0}{\pa\omg_2} = \f\pi2\left[\f{n\omg_1^3}{3\pi^2\hbar\var_0 c^3}\right]\left[\f{n\omg_2^3}{3\pi^2\hbar\var_0 c^3}
    \right]M_{12}^2 \te{ with all } \psi\te{'s}=1,\cr
    \f{\pa\Gamma^\te{TPSE}_{\omg_{1c},\omg_{2c}}}{\pa\omg_2} = \f\pi2\left[\f{2Q_1\phi_1}{\pi\hbar n^2\var_0V_1}\right]\left[\f{2Q_2\phi_2}{\pi\hbar n^2\var_0V_2}\right]
    M_{12}^2
\end{align}
with the energy conservation condition $\omg_1+\omg_2=\omg_d$.

Physically, TPSE occurs as the result of the interaction between the
excited QD and the vacuum states of the electromagnetic field with
modes frequencies $\omg_1$ and $\omg_2$. Since there is a series of
$(\omg_1,\omg_2)$ satisfying the energy conservation condition, the
TPSE spectrum of a single QD is broad. Furthermore, the TPSE rate is
slower than the one-photon emission because virtual states are
required for it. The combination of broad spectrum and low
transition rate results in an overall low signal intensity, and
therefore TPSE is generally difficult to detect.

However, the TPSE rate is enhanced when the QD is placed inside a
cavity, because the cavity modifies the photon density of states in
free space into a Lorentzian distribution and also localizes field
into small volume which leads to stronger interaction. This rate
enhancement is similar to the one-photon spontaneous emission rate
enhancement discovered by Purcell\cite{article:pur46}. The
two-photon rate enhancement in a double-mode cavity relative to the
bulk medium can be seen by taking the ratio of Eqs.~\eqref{eqn:all}:
\begin{align*}
    \f{\Gamma^\te{TPSE}_{\omg_{1c}, \omg_{2c}}}{\Gamma^\te{TPSE}_0} &= F_1F_2 =
    \left[\f{3}{4\pi^2}\left(\f{\lambda_1}n\right)^3\f{Q_1\phi_1}{V_1}\right]\left[\f{3}{4\pi^2}\left(\f{\lambda_2}n\right)^3\f{Q_2\phi_2}{V_2}\right],
\end{align*}
where $F_1$ ($F_2$) and $\lambda_1$ ($\lambda_2$) are the Purcell
factor and free-space wavelength for frequency $\omg_1$ ($\omg_2$),
respectively. It should be noted that this expression corresponds to
the maximum enhancement, assuming the QD is located at the field
maxima for both modes (i.e. all $\psi$'s$=1$).

Experimentally, single-mode cavities with $Q$'s exceeding $10^4$ and
mode volumes below cubic optical wavelength have been demonstrated,
and doubly resonant cavities with similar $Q$'s and mode volumes are
possible\cite{article:vuc02}. Therefore, we expect the QD TPSE to be
increased by up to $4$ orders of magnitude in single-mode cavities
and by $8$ orders of magnitude in double-mode cavities. Second, in
photonic crystal cavities only the emission rates with frequency
pair $(\omg_1,\omg_2)$ matching the cavity frequencies
$(\omg_{1c},\omg_{2c})$ are enhanced while emission rates with other
frequency pairs are greatly suppressed as a result of the reduction
in photon density of states relative to the bulk medium. Thus, the
cavities offer good controls over the frequencies of emitted
photons. Third, inside a double-mode cavity with degenerate
polarizations for both modes, the two emitted photons are entangled
in polarization.

Similarly, we can address the problem of cavity enhanced TPA
(Fig.~\ref{fig:two_photon}c). For bulk, combining $n^2\var_0|\tb
E_1|^2V_1 = N_1\hbar\omg_1$ and $2n\var_0c|\tb E_1|^2A_1=P_1$, we
obtain $N_1\hbar\omg_1= P_1V_1n/(2cA_1)$, where $A_1$ is the laser
beam spot radius. For laser coupled to cavity mode,
$N_1\hbar\omg_1=\eta_1P_1Q_1\phi_1/\omg_1$. Similar expressions hold
for $\omg_2$. By substituting expressions for $N_i\hbar\omg_i$ into
Eqn.~\eqref{eqn:TPSE}, the TPA rates in bulk and double-mode cavity
are, respectively:
\begin{align*}
    \gamma^\te{TPA}_0 &=\f\pi2\left[\f{P_1}{2\hbar^2n\var_0cA_1}\right]\left[\f{P_2}{2\hbar^2n\var_0
    cA_2}\right]M_{12}^2\delta(\omg_d-\omg_1-\omg_2),\cr
    %\Gamma_{\omg_{1c}} &=\f\pi2\left[\f{\eta_1P_1Q_1}{\pi\hbar^2\omg_1n^2\var_0V_1}\phi_1\right]\left[\f{P_2}{2\hbar^2n\var_0
    %cA_2}\right]M_{12}^2\delta(\omg_d-\omg_1-\omg_2), \te{ and}\cr
    \gamma^\te{TPA}_{\omg_{1c},\omg_{2c}} &= \f\pi2 \left[\f{\eta_1P_1Q_1\phi_1}{2\hbar^2\omg_1n^2\var_0V_1}\right]\left[\f{\eta_2P_2Q_2\phi_2}{2\hbar^2\omg_2n^2\var_0V_2}\right]
    M_{12}^2\delta(\omg_d-\omg_1-\omg_2).
\end{align*}
In the nondegenerate TPA, the absorption rate has a linear
dependence on $P_1$ and $P_2$. Similar to the TPSE rate, the TPA
rate is enhanced because the cavity modifies the photon density of
states. For the same excitation laser power and laser frequencies
resonant with cavity mode frequencies, the rate enhancement is
$\gamma^\te{TPA}_{\omg_{1c},\omg_{2c}}/\gamma^\te{TPA}_0 = G_1G_2$,
where $G_i = \eta_iQ_iA_i\lambda_i/(\pi V_in)$.

Two-photon excitation of QDs holds several advantages over
conventional one-photon excitation. By tuning $\omg_1$ and $\omg_2$
to match $\omg_1+\omg_2=\omg_d$, we can coherently excite a QD and
therefore create a source of indistinguishable single photons on
demand. This is because we eliminate a $10$-$30$ ps time jitter in
emitted single photons\cite{article:vuc06}. Two-photon excitation
also presents a solution to the separation of signal and probe
photons, because the excitation and emission wavelengths are
different. Finally, TPA offers a convenient tool to coherently
excite and manipulate a QD, which is of interest for quantum
information processing\cite{article:eng09}.

With similar calculations, we derive the two photon stimulated
emission (TPSTE) rate in a double-mode cavity with stimulation by a
laser beam with power $P_2$ and frequency $\omg_2$, whose coupling
efficiency to the cavity mode is denoted as $\eta_2$,
\begin{equation}\label{eqn:TPSTE}
    \Gamma^\te{TPSTE}_{\omg_{1c},\omg_{2c}} = \f\pi2\left[\f{2Q_1\phi_1}{\pi\hbar n^2\var_0V_1}\right]\left[\f{\eta_2P_2Q_2\phi_2}{2\hbar^2\omg_2n^2\var_0V_2}\right]
    M_{12}^2.
\end{equation}
Fig.~\ref{fig:two_photon}b shows the stimulated emission of a photon
of frequency $\omg_2$ and the spontaneous emission of a photon of
frequency $\omg_1$. Eqn.~\eqref{eqn:TPSTE} shows that the TPSTE rate
is linearly dependent on the stimulation light power and the rate is
increased by a factor of $\eta_2P_2\pi/(4\hbar\omg_2)$ compared to
the TPSE rate. This additional enhancement makes the QD an selective
single photon emitter at $\omg_1$.

\begin{figure}
\includegraphics[scale=1.05]{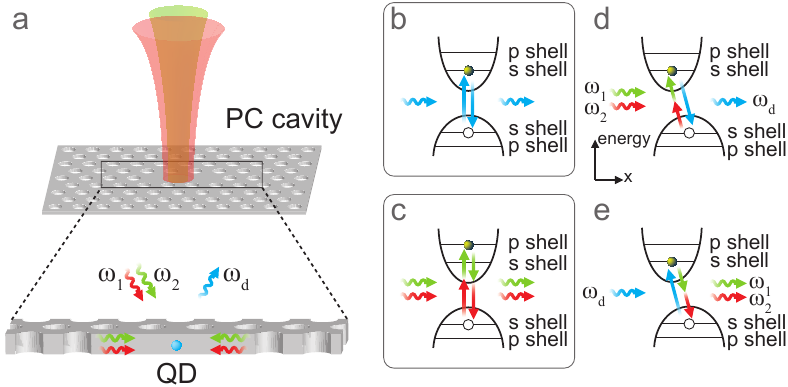}
\caption{\label{fig:setup} (Color online) (a) Setup for enhanced
two-photon absorption experiment: two laser beams with frequencies
$\omg_1$ and $\omg_2$ are injected to a double-mode cavity. (b) and
(c) Parity selection rules dictate that one-photon \textit{s}
shell-\textit{s} shell transitions are allowed but two-photon
\textit{s} shell-\textit{s} shell transitions are forbidden.
However, two-photon \textit{s} shell-\textit{p} shell transitions
are permitted. The \textit{p} shell is approximately $10$ meV above
the \textit{s} shell in InAs QDs\cite{book:mic03}. (d) Enhanced
two-photon absorption and single-photon generation in a QD with a DC
lateral electric field applied to it along the $x$ direction. The
lateral electric field shifts apart the hole valence band and the
electron conduction band and therefore breaks the parity selection
rule. (e) Enhanced two-photon spontaneous emission in a QD as a
reverse process of two-photon absorption. The one-photon excited
electron recombines with the hole and generates entangled photon
pairs at ($1550$, $2300$) nm via two-photon spontaneous emission.}
\end{figure}

We propose to experimentally demonstrate enhanced TPA and TPE in
InAs QDs embedded in two-dimensional GaAs double-mode PC cavities
(Fig.~\ref{fig:setup}a). Since the QDs have ground and excited state
splitting ranging from $900$ nm to $950$ nm, we select two photon
wavelengths as $\lambda_1=1550$ nm and $\lambda_2=2300$ nm,
conveniently coinciding with the telecom band (suitable for
propagation down the fiber) and emission from existing GaSb lasers,
respectively. To generate indistinguishable single photons on
demand, we excite a QD with TPA and collect single photons emitted
at $\lambda_d = 926$ nm (Fig.~\ref{fig:setup}d). To generate
entangled photon pairs, we excite a QD with one-photon process and
collect photon pairs at ($1550$, $2300$) nm via TPSE
(Fig.~\ref{fig:setup}e). In addition, when we stimulate this excited
QD with a $2300$ nm laser beam, we can generate $1550$ nm photons on
demand via TPSTE.

An examination of $M_{12}$ reveals that two-photon transitions have
opposite parity selection rules relative to one-photon transitions.
In one-photon transition, the ground and excited states need to have
opposite parity so that $|\tb d_{ge}|$ is nonzero. On the other
hand, a two-photon transition requires the ground and excited states
having the same parity for $|\tb d_{gk}||\tb d_{ke}|$ to be nonzero.
This selection rule discrepancy implies that two-photon transitions
cannot connect states that have one-photon transitions and vice
versa. For a QD in the effective mass approximation, its energy
eigenfunctions can be written as a product of an periodic function
$u(\tb r)$ and an envelope function $f(\tb r)$, where $f(\tb r)$
satisfies the Schr\"{o}dinger equation. Since $u(\tb r)$ is even for
the valence band states but odd for the conduction band states, an
interband one-photon transition is allowed when its initial and
final state envelop functions have the same parity (i.e. \textit{s}
shell-\textit{s} shell and \textit{p} shell-\textit{p} shell
transitions, as shown in Fig.~\ref{fig:setup}b)\cite{book:mic03},
while an interband two-photon transitions is allowed when its
initial and final state envelope functions have the opposite parity
(i.e. \textit{s} shell-\textit{p} shell transition and vice versa,
as shown in Fig.~\ref{fig:setup}c).

To solve this problem, we propose the application of lateral
electric fields to break the parity symmetry in wavefunctions.
Recent experiments have demonstrated the Stark shift of QD
transitions in bulk\cite{article:rei08}, as well as in photonic
crystal cavities\cite{article:far09,article:lau09}. Following
Refs.~\cite{article:kor09} and~\cite{article:rei08}, we model the QD
as a particle in a finite well along its growth axis and a
two-dimensional harmonic oscillator perpendicular to its growth
axis. We denote the electron (hole) effective mass and oscillator
frequency as $m_e^*$ and $\omg_e$ ($m_h^*$ and $\omg_h$). For the
one-photon \textit{s} shell-\textit{s} shell transition, the lateral
electric field $\mcal E$ reduces this transition dipole moment with
expression $\tb d_{ge} = e\tb r_{cv}\exp[-(\Delta x)^2/(4l_e^2)]$,
where $\tb r_{cv}=\int_Vd\tb r\ u_c(\tb r)\tb r u_v(\tb r)$ is the
transition moment between the valence band and the conduction band
integrated over an unit cell, $\Delta x = e\mcal
E[(m_e^*\omg_e^2)^{-1}+(m_h^*\omg_h^2)^{-1}]$ is the separation of
the electron and hole wavefunction centers, and
$l_e=l_h=\sqrt{\hbar/(2m_e^*\omg_e)}$ is the oscillator length. For
the two-photon \textit{s} shell-\textit{s} shell transition, the
predominant intermediate states are the conduction and valence
\textit{p} shell states. Consider the case of $|k\ra=$ conduction
\textit{p} shell, we have $|\tb d_{gk}| = e|\tb r_{cv}|(\Delta
x/l_e)\exp[-(\Delta x)^2/(4l_e^2)]$, $|\tb d_{ke}| = el_e$, and
therefore $|\tb d_{gk}||\tb d_{ke}| = e^2|\tb r_{cv}|\Delta
x\exp[-(\Delta x)^2/(4l_e^2)]$. The case of $|k\ra=$ valence
\textit{p} shell gives the same product of transition dipole
moments.

\begin{figure}
\includegraphics[scale=.3]{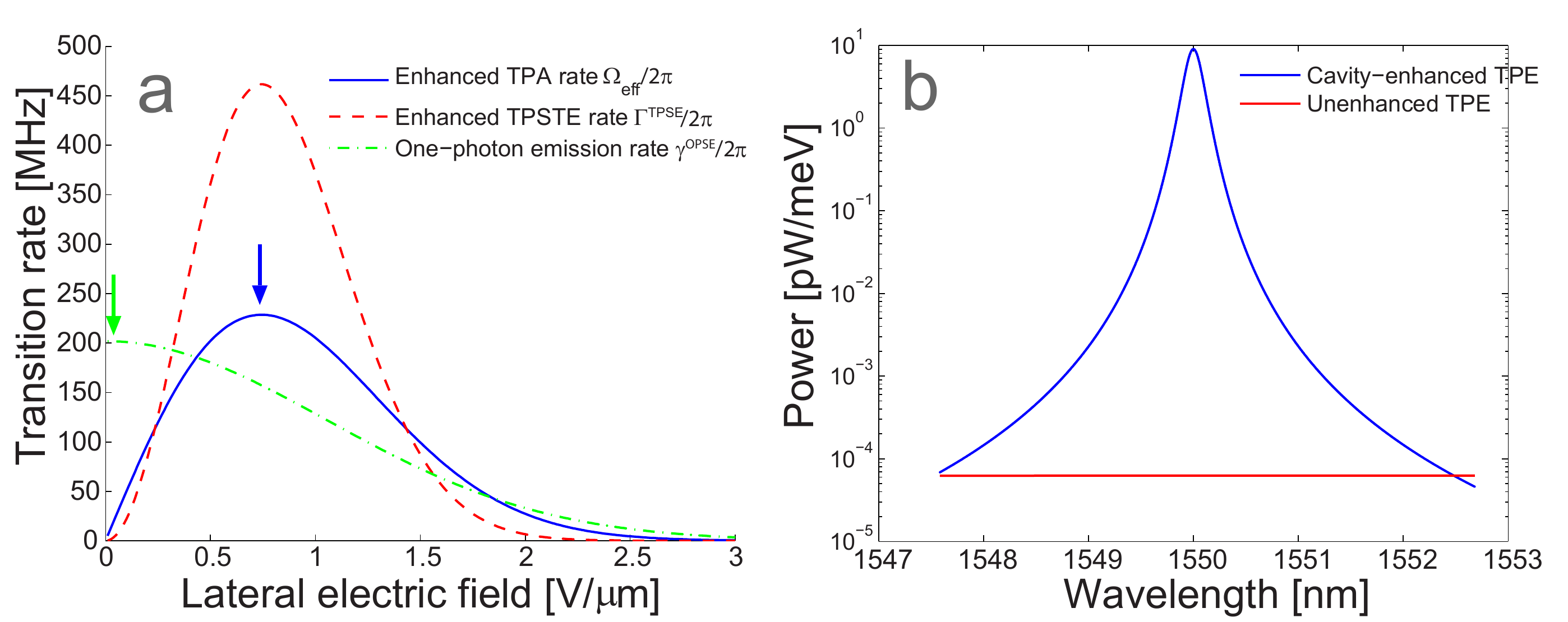}
\caption{\label{fig:TPA} (Color online) (a) Cavity-enhanced
two-photon absorption effective Rabi rate (blue solid line),
enhanced two-photon stimulated emission rate (red dashed line), and
one-photon spontaneous emission rate (green dash-dot line) as a
function of lateral electric field strength. All of the transitions
are \textit{s} shell-\textit{s} shell. The simulation parameters are
shown in text. (b) Cavity-enhanced two-photon spontaneous emission
power (blue) and unenhanced emission power (red) of $1550$ nm
photons. Pairs of photons of $1550$ nm and $2300$ nm wavelength are
emitted from the excited QD.}
\end{figure}

Fig.~\ref{fig:TPA}a shows the enhanced TPA effective Rabi rate
$\Omg_\te{eff}/2\pi$ (Eqn.~\eqref{eqn:nondegenerate_transition}),
enhanced TPSE rate $\Gamma^\te{TPSTE}_{\omg_{1c},\omg_{2c}}/2\pi$
(Eqn.~\eqref{eqn:TPSTE}), and one-photon spontaneous emission rate
$\gamma^\te{OPSE}/2\pi$ as a function of lateral electric field. The
simulation parameters are $m_h^* = 2m_e^* = 0.11m_0$ ($m_0$ is the
electron rest mass), $\hbar\omg_e=2\hbar\omg_h=12$
meV\cite{article:kor09}, $|\tb r_{cv}|=0.6$ nm\cite{article:eli00},
$\eta_1=\eta_2=2\%$, $Q_1=Q_2=5000$, $V_i=(\lambda_i/n)^3$,
$\phi$'s$=1$, and $\psi$'s$=1$. $P_1=P_2=12\ \mu$W for enhanced TPA
and $P_2=100\ \mu$W for enhanced TPSTE. The assumed low incoupling
efficiencies of $2\%$ are taken from the experiments where the input
laser beam is coupled to the cavity in the direction perpendicular
to the chip\cite{article:eng07,article:far08,article:fus08}.
However, it must be pointed out that PC cavities coupled to fiber
tapers or PC waveguide can achieve a coupling efficiency of
$70$-$90\%$\cite{article:hwa05,article:far07} and therefore lower
the needed external excitation power. For enhanced TPA and
subsequent single photon generation (Fig.~\ref{fig:setup}d), we
operate in the region where $\mcal E<0.5$ V/$\mu$m with
$\Omg_\te{eff}<\gamma^\te{OPSE}$ to prevent Rabi oscillations
between the ground and excited states. The electric field of $1$
V/$\mu$m strength only shifts the \textit{s} shell-\textit{s} shell
transition by approximately $0.1$ nm\cite{article:rei08}, and
therefore the QD is still on resonance with the double-mode cavity.
In the case that the cavity also possesses a resonance mode at $926$
nm, the single photon generation rate will be further enhanced by
the Purcell factor at this mode. For enhanced TPSTE used to generate
$1550$ nm single photons, we operate in the region such that
$\Gamma^\te{TPSTE}_{\omg_{1c},\omg_{2c}}>\gamma^\te{OPSE}$ to
prevent $926$ nm one-photon spontaneous emission of the excited QD.
This implies that $\mcal E>0.3$ V/$\mu$m.

In addition to the application of a static electric field, a
modulation of electric field can change the parity selection rules
dynamically. For example, we keep the electric field on at
approximately $0.75$ V/$\mu$m to enable the \textit{s}
shell-\textit{s} shell excitation via TPA (Fig.~\ref{fig:TPA}a blue
arrow), but then turn off the field to permit a rapid single photon
emission (Fig.~\ref{fig:TPA}a green arrow). This way we can permit
both an efficient excitation and an efficient photon emission,
because this approach allows for maximum transition rates in both
steps. For an estimation of the modulation speed, we need to switch
the electric field on a timescale that is faster than the
recombination rate for an excited QD, which is $100$ ps-$1$ ns.
Today's commercial function generators can easily achieve this
speed.

Fig.~\ref{fig:TPA}b shows the power of $1550$ nm photons emitted by
enhanced two-photon spontaneous emission. The figure shows that the
peak enhanced TPSE power from a single QD is on the same order as
the power from a $200$-$\mu$m-thick GaAs sample with carrier
concentration of $1.2\times10^{18}$ cm$^{-3}$\cite{article:hay08}.
%\section{Conclusion}

In conclusion, we have derived the expressions for enhancement of
two-photon spontaneous emission, two-photon stimulated emission, and
two-photon absorption from a single QD in an optical nanocavity. Up
to $8$ orders of magnitude in rate enhancement is possible. We have
also discussed how this could be applied to several important
challenges in quantum information science. In particular, enhanced
two-photon absorption allows coherent excitation of quantum dots and
eliminates jitter, thus enabling generation of indistinguishable
single photons on demand; enhanced two-photon spontaneous emission
generates entangled photon pairs in polarization; enhanced
two-photon stimulated emission further increases the emission rate
and allows single photon generation at telecom wavelengths.

Although we use semiconductor QDs as an illustration for cavity
enhanced TPA and TPE in this paper, we can use the same approach to
excite and manipulate other emitters such as $637$ nm
nitrogen-vacancy centers by laser beams in the $1550$ nm telecom
range. The cavity enhanced TPA and TPE can even enable construction
of hybrid quantum networks containing different emitters just by
changing the frequency of one of the excitation lasers.

\begin{acknowledgments}
The authors gratefully acknowledge financial support provided by
National Science Foundation and the Army Research Office. Z. L. was
supported by the NSF Graduate Fellowship and Stanford Graduate
Fellowship.
\end{acknowledgments}

\bibliography{draft5}

\end{document}